\documentstyle[b98proc,epsfig]{article}

\def\Journal#1#2#3#4{{#1} {\bf #2}, #3 (#4)}

\def\PRL{\em Phys. Rev. Lett.}

\def\PRC{{\em Phys. Rev.} C}

\begin{document}

\title{
STRANGE MAGNETISM
}

\author{T. R. HEMMERT\footnote{Talk given BARYONS 98, 
Bonn, Sept. 22-26,1998. \\
email: th.hemmert@fz-juelich.de}, U.-G. MEI{\ss}NER, S. STEININGER}
\address{FZ J\"ulich, IKP (Theorie), J\"ulich, Germany }

% You may repeat \author and \address if necessary

\maketitle

\abstracts{We present an analytic and parameter-free 
expression for the momentum dependence of the
 strange magnetic form factor of the nucleon $G_M^{(s)}(Q^2)$ 
and its corresponding radius
which has been derived in Heavy Baryon Chiral Perturbation Theory. We also discuss a
 model-independent relation between the isoscalar magnetic and the strange magnetic 
form factors of the nucleon based
 on chiral symmetry and SU(3) only. These limites are used to derive bounds on the 
strange magnetic moment of the
 proton from the recent measurement of $G_M^{(s)}(Q^2=0.1 GeV^2)$ by the SAMPLE 
collaboration.  
}

\section{Introduction}
There has been considerable experimental and theoretical interest
concerning the question: How strange is the nucleon? Despite
tremendous efforts, we have not yet achieved a detailed understanding about
the strength of the various strange operators in the proton. 
A dedicated program at Jefferson
Laboratory preceded by experiments at BATES (MIT) and MAMI (Mainz)
is aimed at measuring the form factors related to the strange vector
current. In fact, the SAMPLE collaboration has recently reported the
first measurement of the strange magnetic moment of the proton~\cite{SAMPLE}. 
To be precise, they give the strange magnetic form factor at a small momentum
transfer, $G_M^{(s)} (q^2=-0.1~{\rm GeV}^2) = +0.23\pm 0.37 \pm 0.15
\pm 0.19\,$nuclear magnetons (n.m.). The rather sizeable error bars document
the difficulty of such type of experiment. On the theoretical side, 
there is as much or even more uncertainty. For example,
the spread of the theoretical predictions for the strange
magnetic moment, $-0.8 \le \mu_p^{(s)} \le 0.5$~n.m.
underlines clearly the abovemade statement. 
In the following we report about
 a parameter--free prediction \cite{letter} for the momentum dependence of
the nucleons' strange magnetic (Sachs) form factor based on the chiral symmetry
of QCD solely. In addition, a leading order
model--independent relation between the strange and the isoscalar
magnetic form factors has been derived, which allows to give an upper bound on the
momentum dependence of $G_M^{(s)} (Q^2)$. These two different results
can then be combined to extract a range for the strange magnetic
moment of the proton from the SAMPLE measurement of the form factor at
low momentum transfer.

\section{Strangeness Vector Current}

The strangeness vector current of the nucleon is defined as
\begin{equation}\label{svc}
\langle N|\;\bar{s}\;\gamma_\mu\; s\;|N \rangle
= \langle N|\;\bar{q}\;\gamma_\mu\;
(\lambda^0/3-\lambda^8/\sqrt{3}) \; q\;| N \rangle 
                                  = (1/3)J_{\mu}^0- (1/\sqrt{3})J_{\mu}^8 \; ,
\end{equation}
with $q=(u,d,s)$ denoting the triplet of the light quark fields and
$\lambda^0 = I\; (\lambda^a)$ 
the unit (the $a=8$ Gell--Mann) SU(3) matrix.
Assuming conservation of  all vector currents, the corresponding singlet and octet 
vector current for a spin--1/2 nucleon can then be written as
\begin{equation}\label{curr}
J_{\mu}^{0,8}
=\bar{u}_N(p^\prime)\left[F_{1}^{(0,8)}(q^2)\gamma_\mu+ F_{2}^{(0,8)}(q^2)
        \frac{i\sigma_{\mu\nu}q^\nu}{2m_N}\right] u_N(p) \; .
\end{equation}
Here, $q_\mu=p^{\prime}_\mu-p_\mu$ corresponds to the four--momentum transfer to the 
nucleon by the external singlet ($v_{\mu}^{(0)}=v_\mu \lambda^0$) and the octet ($
v_{\mu}^{(8)}=v_\mu\lambda^8$) vector source $v_\mu$, respectively.
The strangeness Dirac and Pauli form factors are defined via:
\begin{equation}
F_{1,2}^{(s)}(q^2)= \frac{1}{3}F_{1,2}^{(0)}(q^2)
-\frac{1}{\sqrt{3}}F_{1,2}^{(8)}(q^2) \; ,
\end{equation}
subject to the normalization $F_1^{(s)} (0) = S_B$, with $S_B$ the
strangeness quantum number of the baryon ($S_N = 0$) and
$F_{2}^{(s)} (0) =\kappa_{B}^{(s)}$, with
$\kappa_{B}^{(s)}$ the  (anomalous) strangeness moment. 
In the following we concentrate on the ``magnetic'' strangeness
form factor $G_{M}^{(s)}(q^2)$, which in analogy to the (electro)magnetic
Sachs form factor is defined as
\begin{eqnarray}
G_{M}^{(s)}(q^2)=F_{1}^{(s)}(q^2)+F_{2}^{(s)}(q^2) \label{eq:def1}
\end{eqnarray}
and for which chiral perturbation theory 
(CHPT) gives the most interesting predictions. 

\section{The strange magnetic form factor}

To obtain the complete strange magnetic form factor in ChPT one only has to 
consider the diagrams \cite{letter} where the external singlet/octet source couples 
directly 
to the nucleon as well as the one where the octet source couples to the intermediate
kaon cloud,  the
pion and the $\eta$ cloud do not contribute to this order. 
For the proton ($p$) and the neutron ($n$) one finds
\begin{eqnarray}\label{eq:c}
G_{M}^{(s)}(Q^2) &=& G_{M}^{(s)\;p}(Q^2)\;=\;G_{M}^{(s)\;n}(Q^2) 
\nonumber \\
                &=& \mu_{N}^{(s)}+\frac{\pi m_N M_K}{(4\pi
F_{\pi})^2}\;\frac{2}{3}\left( 5 D^2-6 D F+9 F^2 \right) \, f(Q^2)~,
\label{eq:gms}
\end{eqnarray}
with $Q^2=-q^2$. The strange magnetic moment $\mu_N^{(s)}$ cannot be directly predicted
in ChPT due to the influence of poorly known singlet counterterms \cite{ito,letter}. 
However, to ${\cal O}(p^3)$ in ChPT the momentum dependence is given entirely in terms 
of well-known parameters \cite{letter} and the analytic function
\begin{equation}\label{f2q}
f(Q^2) = -\frac{1}{2} + \frac{4+Q^2/M_K^2}{4\sqrt{Q^2/M_K^2}} 
\arctan \biggl( \frac{\sqrt{Q^2}}{2M_K} \biggr)~. 
\end{equation}
$f(Q^2)$ is shown in Fig.1. For small and moderate $Q^2$,
it rises almost linearly with increasing $Q^2$.
\begin{figure}[ht]
\hspace{1.5in}
\epsfig{file=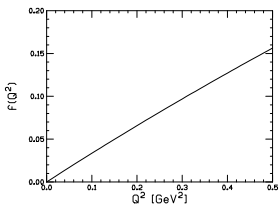,height=2.6cm}
\caption[func]{\protect \small
The function $f(Q^2)$ for small and moderate momentum transfer squared.}
\end{figure}

\section{The isoscalar connection}

An SU(3) analysis of the magnetic isoscalar $(I=0)$ form factor of the nucleon 
$G_{M}^{I=0}(Q^2)$ shows \cite{letter} 
that to ${\cal O}(p^3)$ it can be expressed via the same function $f(Q^2)$ given
in Eq.(\ref{f2q}). We can therefore eliminate $f(Q^2)$ from both expressions and derive 
a {\it model-independent} relation between the isoscalar magnetic form factor 
$G_{M}^{I=0}(q^2)$ of the nucleon and the strange magnetic form factor
\begin{eqnarray}
G_{M}^{(s)}(Q^2) = \mu_{N}^{(s)}+\mu_s-G_{M}^{I=0}(Q^2)+{\cal O}(p^4)\, ,
\label{eq:mi}
\end{eqnarray}
with $\mu_s =0.88\,$n.m. being the isoscalar nucleon magnetic moment.
This relation is {\it exact} to ${\cal O}(p^3)$ in SU(3) heavy baryon CHPT. Possible
corrections in
higher orders can be calculated systematically. This relation again does not constrain
$G_{M}^{(s)}(0)=\mu_{N}^{(s)}$, but makes new predictions on its
$Q^2$-dependence.
Utilizing the {\em empirical} dipole parameterization for $G_{M}^{I=0}(Q^2)$ instead of
the functional form $f(Q^2)$ of Eq.(\ref{f2q}) one now obtains the $Q^2$-dependence
shown in Fig.2 for vanishing $\mu_{N}^{(s)}$. 
Given that there are also non--strange contributions
in the {\em physical} isoscalar magnetic form factor, which will start to manifest at
order $q^4$, we consider Eq.(\ref{eq:mi}) as an {\it upper} bound on the strange
magnetic form factor.
\begin{figure}[ht]
\hspace{1.5in}
\epsfig{file=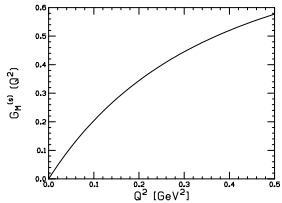,height=2.6cm}
\caption[GMs]{\protect \small
$G_{M}^{(s)}(Q^2)$ derived from the isoscalar magnetic form factor
with $\mu_N^{(s)} =0$.}
\end{figure}

%\begin{figure}[ht]
%\psfig{figure=figure.eps,height=2cm,width=15cm}
%\caption{An invisible nice figure}
%\label{bar}
%\end{figure}

\section{Summary}

In summary, we have derived two novel relations which constrain the
momentum dependence of the strange magnetic form factor in the low
energy region. The first one is based on the observation that to one
loop oder in three flavor ChPT, the strange form
factor picks up a momentum dependence which is free of unknown
coupling constants. The second one rests upon the
observation that the isoscalar magnetic form factor calculated in
SU(3) also acquires a momentum dependence which can be related to the
one of the strange magnetic form factor. 
One can now utilize the $Q^2$--dependence from the two bounds,
Eqs.(\ref{eq:c},\ref{eq:mi}),
to extract the strange magnetic moment from the SAMPLE result for the strange
magnetic form factor. For $Q^2 = 0.1\,$GeV$^2$,
the correction is -0.06 
and -0.20, respectively, i.e. for the mean value
of ref.\cite{SAMPLE} we get
\begin{equation}
\mu_p^{(s)} = 0.03 \ldots 0.18 \, {\rm n.m.} ~,
\end{equation}
which even for the upper value is a sizeable correction. 
Clearly, these numbers should only be considered indicative since
(a) the current experimental errors are bigger than the correction and (b) higher order
corrections to the relations derived here should be worked out. 
Finally, we note that the G0 collaboration at TJNAF will also probe this 
particular range of momentum transfer~\cite{G0}.

We would like to thank the organizers of Baryons98 for providing us with the opportunity
to present this work to the physics community.

% Please use the \Journal macro together with \PLA,\PLB etc. whenever possible!

\end{document}